# A Digital Library for Research Data and Related Information in the Social Sciences

Daniel Hienert, Dagmar Kern, Katarina Boland, Benjamin Zapilko, Peter Mutschke
GESIS – Leibniz Institute for the Social Sciences
Cologne, Germany
{firstname.lastname}@gesis.org

## ABSTRACT

In the social sciences, researchers search for information on the Web, but this is most often distributed on different websites, search portals, digital libraries, data archives, and databases. In this work, we present an integrated search system for social science information that allows finding information around research data in a single digital library. Users can search for research data sets, publications, survey variables, questions from questionnaires, survey instruments, and tools. Information items are linked to each other so that users can see, for example, which publications contain data citations to research data. The integration and linking of different kinds of information increase their visibility so that it is easier for researchers to find information for re-use. In a log-based usage study, we found that users search across different information types, that search sessions contain a high rate of positive signals and that link information is often explored.

## 1 Introduction

Most of today's digital libraries allow researchers to find literature that fits their information needs. However, in empirical sciences such as natural sciences, medicine or the social sciences, researchers are not only looking for literature but also for research data [13, 19], for example, to re-use it for their own research. In the course of Open Science and the FAIR-principles (findable, accessible, interoperable, re-usable) [21], more and more research data sets are archived in specific repositories or are shared through data catalogs for improving the reusability and replicability of research results. While literature and research data sets are usually accessible through different portals, the links between these information items are often missing. Manually curated data catalogs often report references to related publications in the metadata of a research data set. But the other way around, links to a research data set that is mentioned in a paper, are most often not available. So far, researchers have to find connections between literature and research data in a very time-consuming process by querying various search portals and by inspecting the full text of papers for data mentions. In this paper, we introduce a digital library that provides integrated access to research data and publications accompanied by further information such as survey variables, questions from questionnaires and survey instruments and tools. Information items are connected to each other whereby the links are either manually created or automatically extracted by services that find data references in full texts. The linking allows researchers to explore the connections between information items interactively.

In the following, we describe the system and its context by first giving the background information that is needed to understand the domain-specificity of the search system (Section 2). In Section 3, we will give a usage scenario to show what is possible with our portal. We will explain the technical details of the system and how the links between information are generated in Section 4. The portal was built with a strong focus on the user, which will be explained in Section 5. Section 6 then reports on a log-based usage study of the system.

## 2 Background and Related Work

Researchers of all disciplines have common but also different information needs while searching for research data. Gregory et al. [10] provide an overview of data retrieval practices in different domains. Most of the studies focus on why and how researchers search for data sets. In the following, we focus on research data, their retrieval and linking in the field of social science.

## 2.1 Research Data in the Field of Social Science

Research data in the empirical social science are mostly survey data, in which a cross-section of the population is asked about their social background, social facts, their behavior or attitudes towards a specific topic. As the primary data collection and gathering process is very complex and time-consuming, social scientists have a strong desire for re-using data and questionnaires to answer new research questions [5]. This is also reflected in the fact that the identification of data sets for secondary analysis is one basic information need of the scientists [19]. Corti [4] lists re-using, revisiting, reanalyzing and comparing with complementary data sources as factors for re-using qualitative data sets. A number of quality factors such as completeness, accessibility, ease of operation, and credibility [7] can influence the user's satisfaction of reusing data sets.

Different data archives like ICPSR[1], UK Data Service[2], Survey Data Netherlands[3] or GESIS Data Catalogue[4] provide access to such research data. Usually, the stored research data sets consist of different objects: the raw data for statistical analyses, metadata describing the content of the survey data, the questionnaire that is used to collect the data, a method report, and finally the codebook that assigns questions and answer categories to unique, so-called, variables. For example, the fictitious question "What is your attitude towards the European Union?" has the variable "AttitudeEU" which could have the characteristics (1) negative, (2) neutral, or (3) positive. For collecting survey data, researchers can utilize standardized survey instruments for reoccurring questions. For example, the Big Five Inventory is a scale that allows personality assessments. Surveys using the same survey instruments may be easier comparable. Such survey instruments are also provided in specific portals, e.g., CIS[5] (Collection of items and scales).

Kern & Mathiak [16] compare the differences between data set and literature retrieval for empirical German social scientists by conducting a lab study and telephone interviews with 53 participants. They found that quantity and quality of metadata are much more important in data set retrieval than in literature retrieval because going through all the provided documents (codebook, questionnaires, raw data) related to a data set is much more time-consuming than getting a paper and having a quick look at it. Additionally, interlinking is an important factor for the users: first, the direct linking to the survey material such as the documentation, codebook, or questionnaire; second, the linking to primary or secondary literature and third, the linking down to specific variables that have been addressed in a publication [16].

## 2.2 Linking and Retrieving of Social Science Information

On the technical side, the integration, linking, and retrieval of social science information can be achieved with different approaches. One way of interconnecting research data and publications is applied in the "enhanced publications" approach [1]. This approach takes the view of publications as research and dissemination output. One feature of enhanced publications is the association with underlying research data. This can be done by embedding links in the metadata, in a text section, in the references or by making the experiment directly executable within a virtual research environment. For a reliable linkage to data sets a permanent unique identifier from initiatives such as DataCite [3] is needed. Some prototype systems use this approach to show the connections between publications and research data to the end user, e.g., in the context of the OpenAire infrastructure [14] or of Pubmed Central [9].

A more specialized approach is the Data Documentation Initiative (DDI) [20]. It provides an XML-based metadata standard for the description of observational data from the social, behavioral, economic and health sciences. Based on the DDI standard, it is possible to give access to variables and questions from surveys to researchers who want to reuse them in their own studies. A number of data archives and institutions provide information systems for that purpose, e.g., ICPSR, Q-Bank[6], American National Election Studies[7], UK Data Service and Survey Data Netherlands[8]. The interlinking of publication and research data and the deep data model of surveys raise some retrieval problems. One solution proposes the integration of research data and publications based on ontologies and statistical methods that can link down to the questions and variables level [23]. Query expansion on these levels can support retrievability [6], but also human or automatic indexing on all these levels would help [8].

Related work shows that there are mainly individual platforms for accessing either literature, research data sets, survey instruments, or variables/ survey questions. Recently, ICPSR[9] has made publications in their bibliographies searchable and accessible through a new search category. Additionally, they have made the links between publications and research data sets more prominent and provide means for finding variables in research data. Our approach goes three steps further. First, we provide integrated access to information that is available on specific platforms without the physical consolidation of this data. Second, besides the links provided through bibliographies, we generate links automatically by scanning available full texts for the mention of research data and link them to the corresponding research data. Third, we provide further information in the social science such as empirically tested measurement instruments, tool support and best practices in the context of data collection and data analysis that are also linked to research data and publications in our system.

---

[1] https://www.icpsr.umich.edu/icpsrweb/ICPSR/ssvd/index.jsp
[2] http://discover.ukdataservice.ac.uk/variables
[3] http://surveydata.nl
[4] https://dbk.gesis.org/dbksearch/index.asp?db=e
[5] https://www.gesis.org/en/services/study-planning/items-scales/
[6] http://wwwn.cdc.gov/qbank/home.aspx
[7] http://isr-anesweb.isr.umich.edu/_ANESweb/utilities/cbksearch/searchANES.htm
[8] http://surveydata.nl
[9] https://www.icpsr.umich.edu/icpsrweb/

Table 1. Content of the integrated search system.

| Category | Description and Sources | Count |
|---|---|---|
| Research Data | Metadata of studies retrieved from the GESIS data catalogue as well as metadata of studies, harvested from 30 international data providers such as ICPSR, DANS or the UK Data Archive. | 83,699 |
| Questions & Variables[10] | Questions & Variables from research data archived in the GESIS data catalogue. | 103,828 |
| Publications | Open access publications and literature linked to research data. | 94,291 |
| Instruments & Tools | Information about tools for designing surveys and data collection such as pretested response scales, pretested questionnaires or syntax files to conduct analysis in statistical software. | 367 |
| GESIS Webpages | Webpages from the GESIS Homepage including consulting services and trainings for social scientists. | 5,052 |
| GESIS Library | Publications focused on Empirical Social Research and related fields | 134,802 |

## 3 Usage Scenario

The following usage scenario illustrates the possibilities and potentials that our integrated search system offers to social science researchers. *"For a research proposal, Anna Smith is searching for research data and publications on migration to Germany. In particular, she wants to answer the question "How strongly do migrants who have been living in Germany for more than ten years feel connected with Germany?". She uses the integrated search function and first looks at the publications found. One publication attracts her attention, as it appears to be well suited thematically. She notices at first glance the link to the full text, the list of references, and the list of connected research data sets. She downloads the full text of the publication and looks through the list of utilized research data sets. She clicks on one research data set and switches to the corresponding detail view. There, she finds, among other things, free access to the questionnaires, the codebook, and the download link to the data set. Connected to the research data set, she finds a survey instrument that addresses a similar issue she is interested in. In addition, she gets the list of questions and variables used in the study and found a well-suited question that deals with feelings of connection to Germany. As she wants to know if a similar research question to hers is already answered with this data set, she looks at the list of publications that referenced it."* The usage scenario based on results collected in several user studies (cf. Section 5).

## 4 The Integrated Search System

In this section, we give an overview of the integrated search system by outlining the content, the technical architecture, the main functionality, the user interface, and the link infrastructure. The system can be accessed via the following URL: https://search.gesis.org.

### 4.1 Content

At present, the system provides an integrated search across all data collections at GESIS, i.e., metadata on research data, questions from questionnaires, survey variables, instruments and tools for creating surveys or analyzing data, publications as well as GESIS webpages and the GESIS library. The focus of the content is on research data, open access publications, and related information. For research data, first, we give access to 6,267 studies from 1945 to 2018 that are collected in the GESIS Data Archive and made accessible through the GESIS data catalog. Second, we include about 77,432 studies from international players such as ICPSR, DANS (Data Archiving and Networked Services Netherlands) or the UK Data Archive which have been harvested and made accessible in the da|raSearchNet project [18]. This gives the so-far best overview of research data in the field of social sciences. Open access publications come mainly from the SSOAR[11] portal, which is an open access repository for social science researchers to upload their full texts. References from open access texts are extracted, according to [17], to allow users to browse through literature references. Other information types are included because of their relationship to research data, e.g., questions & variables used in surveys of the research data, literature that cites research data or survey instruments that are used for collecting these research data. Table 1 gives an overview of the different categories and their content.

### 4.2 Technical Architecture

The integrated search system consists of the following technical components: 1) the web application, 2) the search engine and 3) the link database (Link-DB) shown in Figure 1. The web application is a JavaScript client software based on FacetView from the Open Knowledge Foundation[12]. It communicates with the search engine through a web interface (API).

---
[10] this category is currently only available in the test system under https://searchtest.gesis.org
[11] https://www.gesis.org/ssoar/home/
[12] https://github.com/okfn/facetview

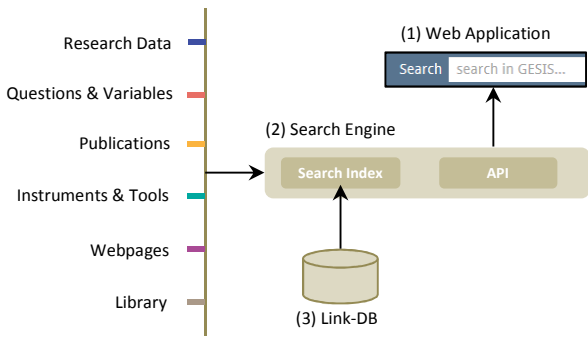

**Figure 1: Architecture of the integrated search system.**

Elasticsearch[13] is used as the underlying search engine. The search index is updated every night with data from 14 different endpoints such as search indexes, and data dumps holding the different information types, links, references and full-text information. The whole indexing process is optimized for speed and takes about one hour.

In order to integrate metadata from different portals into a central search index, a common metadata schema was created based on Dublin Core and enhanced with fields that are specific for each information type. Each data record is also enriched with information from the Link-DB that contains the links between the different information items and will be explained in detail in section 4.4.

### 4.3 Functionality

The web application provides the following core functionality to the end user: An integrated search over research data sets, publications, questions & variables, instruments & tools, GESIS websites, and the GESIS library. If, for example, the user searches for a topic such as "gender roles" the system returns results for all categories. For each category, the numbers of hits are shown (see Figure 2).

The user can switch between the different categories to see the corresponding result lists. If the user clicks, for example, on "Research data", only results in this category are shown (see Figure 2). Here, the search can be narrowed down either by providing more search terms or by choosing from different facets. Each result item is presented with its core metadata, materials (data set, codebook, etc.), full texts and actions such as the possibility to cite the record. A context snippet is shown in which the user search terms have been found, and the search term is highlighted in the metadata or in the snippet. The numbers of links to related information are shown as buttons under each result item.

By clicking on a title, the user can switch to the detailed view of a record. Here, all metadata fields of the record are presented in detail (see Figure 3). Information linked to this record is presented in separate boxes for each information type which can be unfolded to see the list of linked information. Each record can again be clicked and inspected on its detailed view page. This way, the user can browse through related information. Figure 3 shows an example with a detailed view of the "International Social Survey Programme: Family and Changing Gender Roles IV - ISSP 2012". Related publications and applied instruments are shown in separate boxes. The latter one is unfolded and shows links to scales that have been used in the study.

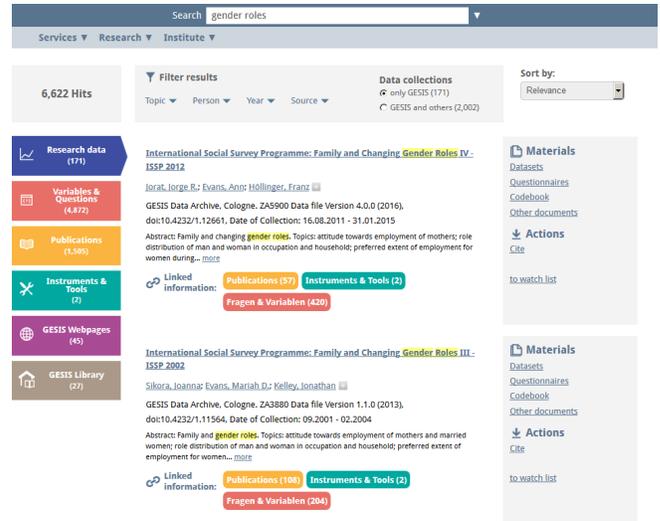

**Figure 2: View of the result list for the search query "gender roles".**

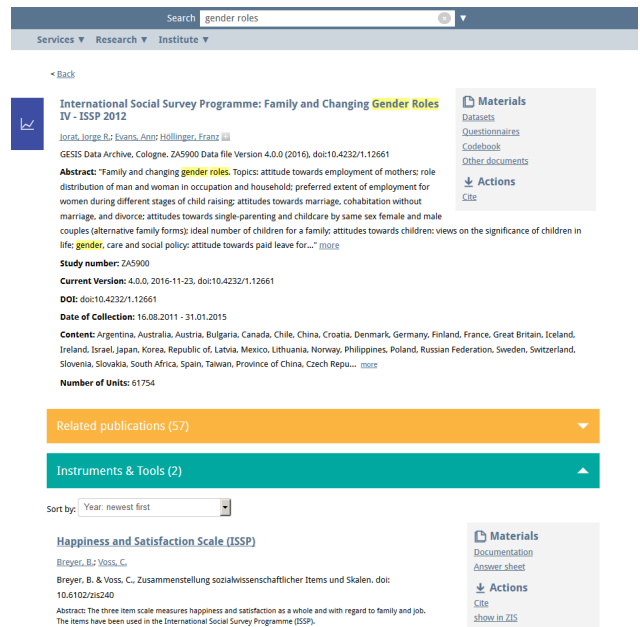

**Figure 3: Detailed view of a record with linked information and "Applied instruments & tools" unfolded.**

---
[13] https://www.elastic.co/products/elasticsearch

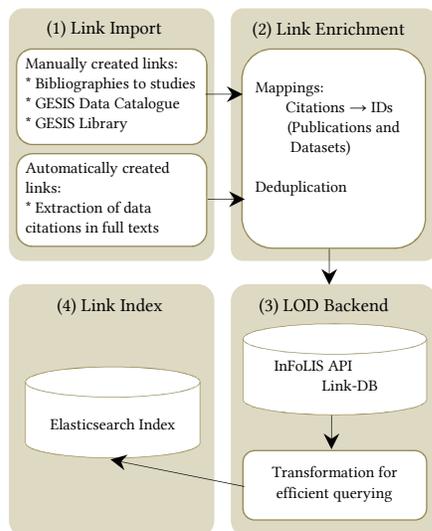

**Figure 4: Workflow of the link infrastructure.**

## 4.4 Link Infrastructure

All links between information items that are shown in the integrated search system are stored in the link infrastructure. In the following section, we give a brief overview of how these links are collected, enriched and maintained. The workflow for importing all data and building up the Link-DB is depicted in Figure 4. More detailed technical information about the link infrastructure can be found in [22].

*(1) Link Import*: The links between information entities in the integrated search are drawn from a variety of sources. Links are either manually curated or automatically generated. Manually curated links are provided by bibliographies containing all known publications related to a given research data set or to given survey instruments, by the GESIS data catalog that provides links between data sets and primary literature, and by GESIS library employees who create links by discovering references to research data set in publications. Automatically generated links are provided by utilizing the InfoLink tool[14] [2] which identifies automatically the mentioning of research data sets in publications that are available as full texts.

*(2) Link Enrichment:* Only a share of the imported links contains unique identifications (e.g., DOIs) to link them unambiguously to each other. For the majority of extracted data citations from full text, there is no explicit ID provided e.g., "we used the ISSP 2010". In this case, the registration agency for social and economic data (da|ra[15]) is queried to find the right data set. It can be the case that there are more than one research data set that would fit. In this case, we provide links to all research data sets in question sometimes with different confidence values (on a scale from 0 to 1, whereby 1 is the highest) according to the degree of the match. In the interface, we label the data with lower confidence than 1 as "mentioned data sets" and the others as "used data set" to make the differences clear. Additionally, we provide in most cases the extracted text passage for the "mentioned data sets".

To set unique links to publications, a database of more than 9 million literature records [12] is queried with mentioned references. For references that cannot be mapped to any existing entity, a new entry based on the available metadata is created and stored in an internal literature pool with a distinct ID. The source data may contain duplicates of information items and/or links between them, both within a source and across sources. For this reason, information items are deduplicated during import using their IDs and basic metadata. When duplicate entries are found, their metadata and all outgoing and incoming links are merged.

*(3) LOD Backend:* The link infrastructure reuses the InFoLiS infrastructure[16] consisting of a Node.js based API backend which provides RESTful web services and a LOD (linked open data) representation of the data. The data is stored in a MongoDB in an RDF-based data model. Storing the data in graphs allows easy representation of the links between items and storing additional, necessary information like provenance information. This is highly important to comprehend how and on which basis links have been generated.

*(4) Link Index:* To supply efficient access to all data, the links are transformed and indexed using Elasticsearch. Table 2 gives an overview of linking frequencies between different information types that are contained in the link index so far.

**Table 2: Number of links between different information types.**

| Links between information types | Count |
| --- | --- |
| Research Data ↔ Publication (Manually) | 25,940 |
| Research Data ↔ Publication (Automatically) | 24,791 |
| Research Data ↔ Questions & Variables | 103,833 |
| Research Data ↔ Instruments & Tools | 89 |
| Publication ↔ Instruments & Tools | 2,692 |

## 5 User-Centered Design Process

The integrated search system is developed following the user-centered design process according to ISO 9241-201:2010[17]. Table 3 provides an overview of the performed user studies to understand the context of use, to specify user requirements, to evaluate design decisions and usability. In total, we conducted two observational studies, telephone interviews, a diary study, several iterations of click dummy tests and three comprehensive usability studies. All participants were social scientists with different educational backgrounds ranging from Bachelor degree to habilitation. A few of them worked at our institute but were not involved in the development of the integrated search system. Most of the participants were external social science researchers recruited by mailing lists, the IIRpanel [18], or poster announcement at the University of Cologne, Düsseldorf, and Mannheim.

---
[14] https://github.com/infolis/infoLink
[15] https://www.da-ra.de/en/home/
[16] https://github.com/infolis
[17] https://www.iso.org/standard/52075.html

**Table 3: User studies performed during the user-centered development process of the integrated search system.**

| Title | Participants | Procedure | Purpose |
|---|---|---|---|
| Observational study - literature search | 15 social science researchers (7 female, 8 male) | 10min literature search in Sowiport interviews | Understanding the context of use, identifying typical tasks |
| Observational study - research data search | 7 social science researchers (3 female, 8 male) | 10min data set search in the GESIS data catalogue, interviews | Understanding the context of use, identifying typical tasks |
| Telephone interviews – Research data search | 46 GESIS data catalogue users (11 female, 35 male) | Interviews on requirements and suggestions regarding data search | Specifying user requirements |
| Diary study | 12 social science researchers (5 female, 7 male) | Protocolling information needs over a period of 2 weeks | Studying information needs, specifying use cases |
| Click dummy tests | 5 participants per iteration | Iterative tests of different interaction concepts with prototypes | Evaluating design decisions (e.g. menu structure, result list presentation) |
| 1. Usability study | 18 social science researchers (7 female, 11 male) | Performing a task with the high-fidelity prototype + free exploration, participants were encouraged to think-aloud | Evaluating usability, linking concept |
| 2. Usability study | 4 iterations with 3 social science researchers each (5 female, 7 male) | Performing a task with the developed integrated search system + free exploration, participants were encouraged to think-aloud | Evaluate usability |
| 3. Usability study | 2 iterations with 3 social science researchers each (2 female, 4 male) | Free exploration, participants were encouraged to think-aloud | Evaluating usability before product lunch |

The user studies to understand the context of use and to specify user requirements show the diversity of information needs in the field of Social Science. We categorized the collected information needs into the following categories:

(1) Literature: e.g., I'm looking for full texts.
(2) Research data: e.g., I'm looking for a research data set mentioned in a paper.
(3) Variables in research data: e.g., I'm looking at a specific variable.
(4) Measurement instruments: e.g., I'm looking for scales for a specific question.
(5) Support for data analysis: e.g., I'm looking for information on a specific analysis method.
(6) Support for data collection: e.g., I'm looking on information on survey methods.
(7) Software support: I'm looking for Stata tutorials.
(8) Networking/cooperation: e.g., I'm looking for project partners.
(9) Illustrative material: e.g., I'm looking for videos or images.

In the search interface, we combined the support needs (4-7) into the category "Instruments and Tools". Literature (1), research data (2), and variables and questions (3) got their own categories because of their expressed importance and the comprehensive data in each category. So far networking and illustrative material are not included in the integrated search system. Based on these study results, we created the usage scenario presented in Section 3. More details about the collected information needs can be found in [15].

In the usability studies, participants had to explore the integrated search system either with a given scenario or/and with a self-selected search topic in mind. They provided feedback through thinking-aloud while performing the task and by answering interviewer questions. In doing so, we found a couple of usability problems to fix. However, we also got a lot of positive feedback on the interaction and linking concept. The linkage between information items is one of the core functionality of the integrated search portal and was specially addressed in the first usability study. In an evaluation scenario, users had to experience the capabilities of the provided links for exploring connected research information first by a giving scenario and afterward with their own information need. Regarding the linking concept, we asked 17 of the 18 participants for their assessments on the usefulness of the linked information. 12 participants rated the links as very useful, four as useful and one found it neither useful nor not useful (collected through a Likert-scale ranging from 1="not useful at all" to 5="very useful", mean = 4.65, SD=0.5). More evaluation results regarding the basic linking can be found in [22].

Besides the most positive feedback, we identified some challenges. When participants could freely explore the integrated search function, we observed that after following a couple of links, participants had problems to get back to their starting point. We are currently investigating visualization techniques that should help to keep the overview and prevent the feeling of getting lost in the system. Furthermore, some of the participants felt misdirected when they found out that the provided links represent all connections the information item has and not only those that are related to their individual information need. In this case, appropriate labeling and ranking of the linked information items according to the information need might be helpful.

---
[18] www.gesis.org/IIRpanel

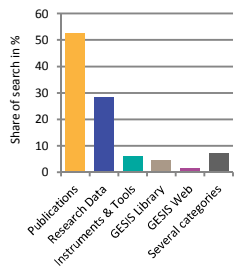 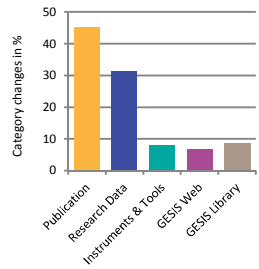 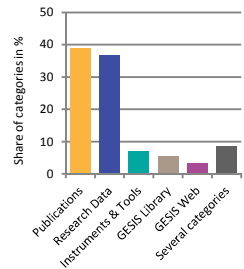 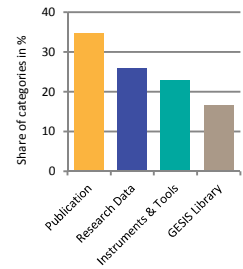

**Figure 5a: Share of searches over different categories**  **Figure 5b: Share of changes to target category**  **Figure 5c: Share of categories starting with search**  **Figure 5d: Share of categories starting with view_record**

The iterative development together with social science researchers has enabled us to design a system that matches the information needs and requirements of our target group. The followed user-centered design process did not stop after the launch of the integrated search function in August 2017. We constantly add new functions that are also developed according to the principles of user-centered design.

## 6 Evaluation of Usage

In this section, we report about results from a log-based usage study of the integrated search. This gives us a first impression on how the different information categories are used, if search sessions are successful and how the linkage between information items are explored.

User interactions in the integrated search are logged and saved to a database; 43 different actions are logged. The most basic user actions are a search conducted by a user (*search*) and a visit of the detailed view of a record (*view_record*). Other actions are changes of searches (e.g., paging), clicks on links (e.g., go to a specialized portal) or the use of linked information.

For this study, we use an evaluation dataset of about three months from 10[th] August 2018 until 16[th] November 2018. It contains 17,143 user sessions with 222,495 user actions. A user session is on average 6:57 minutes long and contains 12.97 user actions. The most frequently occurring actions in the evaluation dataset are the search action (56.48%) and the detailed view action (11.13%).

### 6.1 Categories

First, we have a look at the usage in different categories. Note that the *questions & variables* category is not part of this study, because it is yet only available in the test system. Figure 5a shows how distinct search actions are distributed over different categories. Most searches have been conducted in publications (52.48%) and in research data (28.12%), then follows on a lower level the categories instruments & tools (5.94%), GESIS library (4.65%), GESIS web (1.52%) and across all categories (7.27%). However, users do not always remain in the same category; in 36.93% of the sessions, the user changed the category at least once. Figure 5b shows the distribution of target categories. From 12,736 category changes most users changed to publications (45.21%) and research data (31.44%), then follows instruments & tools (8.05%), GESIS web (6.68%) and GESIS library (8.59%).

There are different ways to start the session. Users can either conduct a search directly from the search bar inside the system or go directly to a detailed view of a record by coming from outside the system, e.g., from a search engine, from a link on the Web or by clicking on a bookmark. We count 13,447 sessions (78.44%) in which the first action is a *search*, 3,224 sessions (18.8%) in which the first action is a *view_record* and 472 sessions that start with other action types. Figure 5c shows the share of categories that start with a search action. The categories publication (38.93%) and research data (36.88%) are the main entry points for sessions starting with a search. Sessions that start with a *view_record* action (see Figure 5d) are more evenly distributed among publications (34.74%), research data (25.81%), instruments & tools (22.98%) and GESIS Library (16.47%).

### 6.2 Successful Sessions

We also measured how successful a user session is and how our system supports that based on a method described in [11]. We, therefore, defined a number of signals that show a particular user interest in a record. Table 4 gives an overview of the positive signals in the integrated search. For example, the user can directly download the research data set or the full text. If not available, the full text can also be searched on outside services such as Google Scholar. Users can also export citations of a record for the use in their personal reference manager. Some records have links to their counterpart in a specialized portal where more detailed information and materials can be found. A last group of signals concerns the usage of the linked information section underneath each record.

39.73% of all sessions contain positive signals. Overall, we found 24,556 positive signals with an average of 3.60 (sd: 4.33)

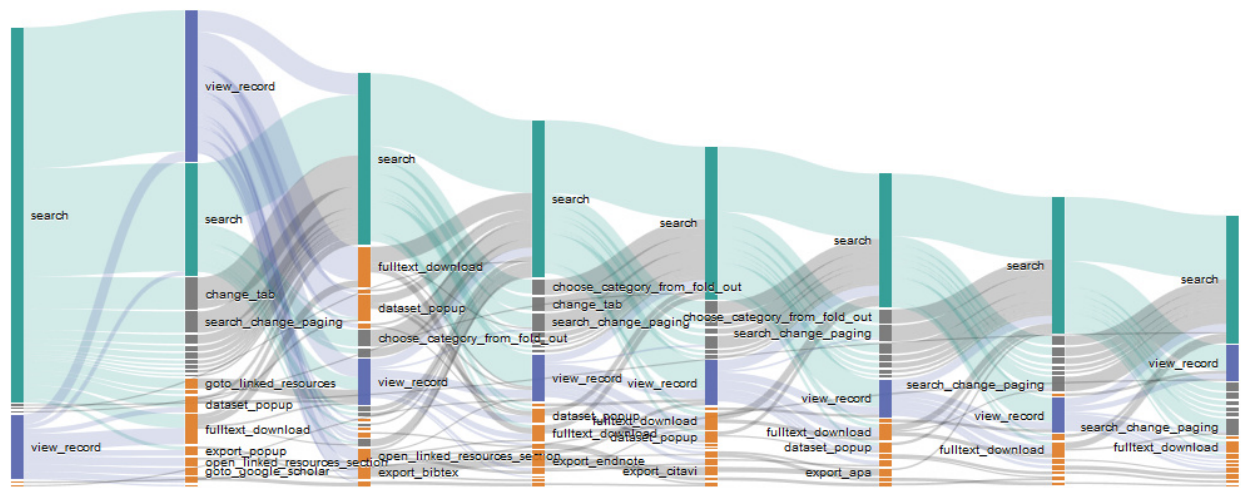

Figure 6: First eight user actions in n=6,812 sessions with positive actions. Search actions in green, view_record actions in blue, positive signals in orange, other actions in grey.

positive actions per session. High shares on positive signals are the direct download of full texts (31.64%), opening the window for research data download (13.81%) and opening the linked resource section in the detailed view (9.47%). Additionally, the export citation functionality seems to be important with a sum of about 20%.

Figure 6 gives an overview of the first eight user actions in all sessions with positive signals (n=6,812). Searches (colored in green) and record views (colored in blue) are the predominant actions in the action paths. In the first step, a search action leads to record views (blue), new searches (green), search modifications such as paging (grey), but also to positive signals (orange). This is because in the hit list users can directly click buttons and links to download the dataset or full text, to export the citation, or to go to the linked resources section. A *view_record* action in the first session step either leads to a new *view_record* action or a *search*, but most often directly to positive signals.

In the following steps, the behavioral patterns repeat. Searches most often lead to new searches, search modifications or record views. However, beginning from step two, most positive signals origin from record views and only rarely directly from the hit list.

### 6.3 Link Views

As the focus of our digital library is on linked information, an important question is if users are able to recognize links between information items in the system. Only a part of all information items are linked (see table 3), so we have to filter the evaluation dataset to all sessions containing at least one record view with links to related objects. This results in 3,846 sessions with 7,778 *view_record_links* actions containing links to related objects. From this possible record views, in 3,422 (44.00%) cases users opened and viewed the link section either by clicking the button to related objects in the hit list or by opening the related objects section directly in the detailed view. Figure 7 shows the first eight actions for the 1,747 sessions in which the link section was opened at least once. Record views with related information (in blue) lead to the opening of the linked resources section (in red). This is then either followed by new searches, new record views, but very often also directly by positive signals such as full text or dataset download (in orange).

Table 4: Positive signals in the integrated search. Shares >5% in bold.

| Signal Group | Signal | Percent |
|---|---|---|
| Research dataset and materials download | dataset_popup | **13.81%** |
| | questionnaire_popup | 4.36% |
| | otherdocs_popup | 1.27% |
| | codebook_popup | 0.77% |
| Full text direct download | fulltext_download | **31.64%** |
| Find the full text outside | goto_google_scholar | **5.48%** |
| | goto_google_books | 0.43% |
| Export citation | export_bibtex | **5.86%** |
| | export_citavi | **5.55%** |
| | export_endnote | **5.48%** |
| | export_popup | 4.52% |
| | export_apa | 4.11% |
| Go to a specialized portal | goto_zis | 0.53% |
| | goto_pretest | 0.06% |
| | goto_survey_guidelines | 0.05% |
| | goto_gml | 0.00% |
| View linked resources | open_linked_resources_section | **9.47%** |
| | goto_linked_resources | 3.62% |
| | click_on_linked_resource | 1.34% |
| | open_linked_resources | 0.84% |

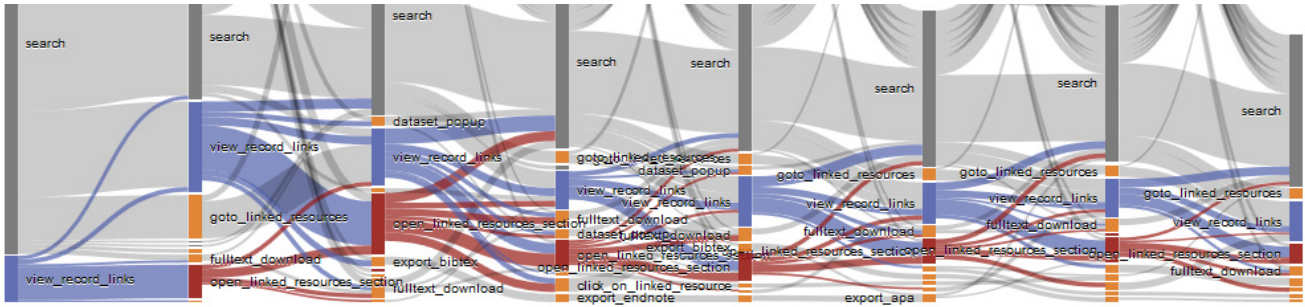

**Figure 7: First eight user actions in n=1,747 sessions in which the user opened the link section (action in red). Record views with related information (view_record_links) in blue and positive signals in orange.**

Figure 8 shows which link paths are viewed by the users. Most links viewed are from research data to publications (28.87%), and vice versa (19.81%), instruments & tools to research data (13.15%) and GESIS Library to research data (12.07%). Research data and publications play an important role on the source and target side. Links from Instruments & Tools and GESIS Library are frequently followed, although they only account for a small part of the total link set.

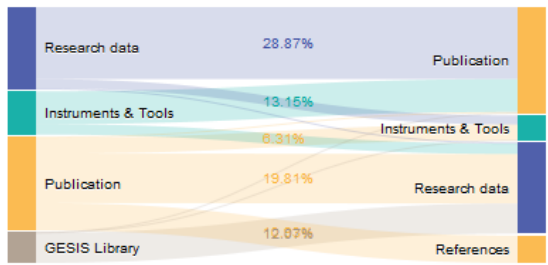

**Figure 8: Direction of links viewed by users for n=3,422 opened link sections**

## 7 Conclusion & Future Work

The digital library presented in this paper provides a central search entry point to the information space of empirical social science. The system highlights links between different information objects, namely, between publications and research data, research data and survey instruments, research data and survey questions, etc. Besides manually curated links, it exploits automatically extracted links. For that, an algorithm identifies citations of survey data in publications, available as full texts, and links them to corresponding research data sets.

Representatives from our target group have been involved in the development process from the beginning on. They emphasized, in particular, the benefit of the linking concept. For example, they appreciate to see directly for a publication which research data set is related to this publication and for a research data set which publications cite or mention this data set. In this way, users are able to find quickly related publications to see how the data in question can be used and what kind of analysis is already performed with them. Our user studies within the frame of a user-centered design approach indicate a high interest of users for an integrated search system and for linked information. However, more research is certainly needed here to develop interaction concepts that support users comprehensively in exploring this linked information.

Complementing that, in a log-based usage study of real-world users we found that:

(1) Users actively search across different categories such as research data, publications or instruments & tools.
(2) A high rate of about 40% of user sessions contains positive signals such as full text or research data set download.
(3) Link sections are viewed with a high rate of 44% where available. These then lead directly to positive signals, or to new searches and record views.

In future work, we intend to extend the granularity of the linking between publications and research data. So far, the system links from a publication to a research data set. However, most often only a couple of often several hundred questions and variables are addressed in a publication. Thus, the benefit of linking would gain a lot if applied on the variable level as well. Zielinski & Mutschke [24] discusses first approaches on how survey variables can be extracted from full texts. Additionally, visualizing, ranking and filtering of link information will be important issues for future work.

## ACKNOWLEDGMENTS

We thank all persons who participated in the development of the integrated search as part of the working group, by giving feedback, by opening their data sources, or by participating in a user study. A number of past or ongoing DFG projects contributed to the development of the integrated search such as SSOAR (#32562673), InFoLIS (#189200501), da|raSearchNet (#189482514), and EXCITE (#293069437).